\documentclass[twocolumn,english,aps,prl,showpacs,superscriptaddress]{revtex4}
\usepackage[T1]{fontenc}
\usepackage[latin1]{inputenc}
\usepackage{graphicx}

\makeatletter
\@ifundefined{textcolor}{}
{%
 \definecolor{BLACK}{gray}{0}
 \definecolor{WHITE}{gray}{1}
 \definecolor{RED}{rgb}{1,0,0}
 \definecolor{GREEN}{rgb}{0,1,0}
 \definecolor{BLUE}{rgb}{0,0,1}
 \definecolor{CYAN}{cmyk}{1,0,0,0}
 \definecolor{MAGENTA}{cmyk}{0,1,0,0}
 \definecolor{YELLOW}{cmyk}{0,0,1,0}
 }

\@ifundefined{definecolor}
 {\usepackage{color}}{}
\@ifundefined{definecolor}
 {\@ifundefined{definecolor}
 {\usepackage{color}}{}
}{}
\@ifundefined{definecolor}
 {\@ifundefined{definecolor}
 {\@ifundefined{definecolor}
 {\usepackage{color}}{}
}{}
}{}
\@ifundefined{definecolor}
 {\@ifundefined{definecolor}
 {\@ifundefined{definecolor}
 {\@ifundefined{definecolor}
 {\usepackage{color}}{}
}{}
}{}
}{}
\makeatother

\makeatother

\usepackage{babel}

\makeatother

\usepackage{babel}

\makeatother

\usepackage{babel}

\makeatother

\usepackage{babel}

\makeatother

\usepackage{babel}

\begin{document}

\title{Quantum Critical Transport Near the Mott Transition}

\author{H. Terletska}

\affiliation{Department of Physics and National High Magnetic Field Laboratory,
Florida State University, Tallahassee, Florida 32306, USA.}

\author{J. Vu\v{c}i\v{c}evi\'{c}}

\affiliation{Scientific Computing Laboratory, Institute of Physics Belgrade, University
of Belgrade, Pregrevica 118, 11080 Belgrade, Serbia.}

\author{D. Tanaskovi\'{c}}

\affiliation{Scientific Computing Laboratory, Institute of Physics Belgrade, University
of Belgrade, Pregrevica 118, 11080 Belgrade, Serbia.}

\author{V. Dobrosavljevi\'{c}}

\affiliation{Department of Physics and National High Magnetic Field Laboratory,
Florida State University, Tallahassee, Florida 32306, USA.}
\begin{abstract}
We perform a systematic study of incoherent transport in the high
temperature crossover region of the half-filled one-band Hubbard model.
We demonstrate that the family of resistivity curves displays characteristic
quantum critical scaling of the form $\rho(T,\delta U)=\rho_{c}(T)f(T/T_{o}(\delta U))$,
with $T_{o}(\delta U)\sim|\delta U|^{z\nu}$, and $\rho_{c}(T)\sim T$.
The corresponding $\beta$-function displays a {}``strong coupling''
form $\beta\sim\ln(\rho_{c}/\rho)$, reflecting the peculiar mirror
symmetry of the scaling curves. This behavior, which is surprisingly
similar to some experimental findings, indicates that Mott quantum
criticality may be acting as the fundamental mechanism behind the
unusual transport phenomena in many systems near the metal-insulator
transition. 
\end{abstract}


\maketitle
Many systems close to the metal-insulator transition (MIT) often display
surprisingly similar transport features in the high temperature regime
\cite{Abrahams2001,Limelette2003,Kagawa2005}. Here, the family of
resistivity curves typically assumes a characteristic {}``fan-shaped''
form (see Fig. 1(a)), reflecting a gradual crossover from metallic
to insulating transport. At the highest temperatures the resistivity
depends only weakly on the control parameter (concentration of charge
carriers \cite{Abrahams2001}, or pressure \cite{Limelette2003,Kagawa2005}),
while as $T$ is lowered, the system seems to {}``make up its mind''
and rapidly converges towards either a metallic or an insulating state.
Since temperature acts as a natural cutoff scale for the metal-insulator
transition, such behavior is precisely what one expects for quantum
criticality \cite{Dobrosavljevic1997}. In some cases \cite{Abrahams2001},
the entire family of curves displays beautiful scaling behavior, with
a remarkable {}``mirror symmetry'' of the relevant scaling functions
\cite{Dobrosavljevic1997}. But under which microscopic conditions
should one expect such scaling phenomenology? What is the corresponding
driving force for the transitions? Despite recent progress, such basic
physics questions remain the subject of much ongoing controversy and
debate.

The phenomenon of disordered-driven Anderson localization of noninteracting
electrons is at present rather well understood based on the scaling
formulation \cite{Abrahams1979}, and is generally viewed as an example
of a $T=0$ quantum phase transition (QPT). On the other hand, a considerable
number of recent experiments \cite{Abrahams2001} provide compelling
evidence that strong correlation effects - some form of Mott localization
- may be the dominant mechanism \cite{Camjayi2008}. Should one expect
similar or very different transport phenomenology in the Mott picture?
Is the paradigm of quantum criticality even a useful language to describe
high temperature transport around the Mott point? These issues are
notoriously difficult to address, because conventional Fermi liquid
concepts simply cannot be utilized in the relevant high temperature
\emph{incoherent} regime. In this Letter we answer this question in
the framework of dynamical mean-field theory (DMFT) \cite{Georges1996},
the only theoretical method that is most reliable precisely at high
temperatures.%
\begin{figure}[t]
 \includegraphics[width=3.3in]{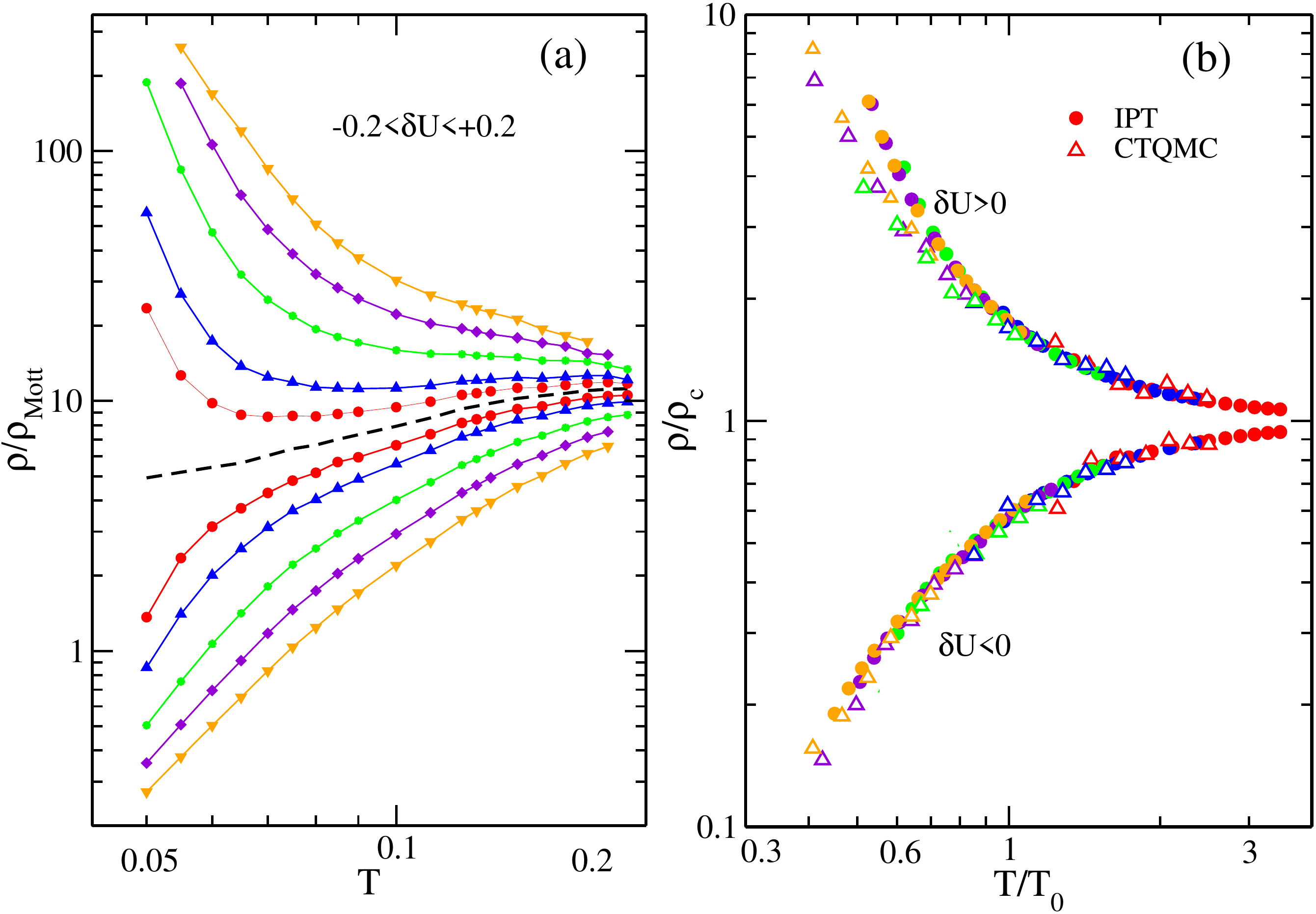} \caption{(Color online) (a) DMFT resistivity curves as function of temperature
along different trajectories $-0.2\le\delta U\le+0.2$ with respect
to the instability line $\delta U=0$ (black dashed line, see the
text). Data are obtained using IPT impurity solver. (b) Resistivity
scaling; essentially identical scaling functions are found from CTQMC
(open symbols) and from IPT (closed symbols)}

\vspace*{-12pt}
 
\end{figure}

\textit{Model} \emph{and DMFT solution}. We consider a single-band
Hubbard model at half-filling \begin{equation}
H=-\sum_{<i,j>\sigma}t_{ij}\left(c_{i\sigma}^{\dagger}c_{j\sigma}+c.c.\right)+\sum_{i}Un_{i\uparrow}n_{i\downarrow},\label{eq:}\end{equation}
 where $c_{i\sigma}^{\dagger}$ and $c_{i\sigma}$ are the electron
creation and annihilation operators, $n_{i\sigma}=c_{i\sigma}^{\dagger}c_{i\sigma}$,
$t_{ij}$ is the hopping amplitude, and $U$ is the repulsion between
two electrons on the same site. We use a semicircular density of states,
and the corresponding half-bandwidth $D$ is set to be our energy
unit. We focus on the paramagnetic DMFT solution, which is formally
exact in the limit of large coordination. Here the Hubbard model maps
onto an effective Anderson impurity model supplemented by a self-consistency
condition \cite{Georges1996}. To solve the DMFT equations we use
the iterated perturbation theory (IPT) \cite{Georges1996} and cross-check
our results with numerically exact continuous time quantum Monte Carlo
(CTQMC) \cite{Werner2006,Haule2007}. We find, in agreement with previous
work \cite{Kotliar2000}, that after appropriate energy rescaling
(see below), the two methods produce qualitatively and even quantitatively
identical results in the incoherent crossover region that we examine.
\begin{figure}[t]
 \includegraphics[width=3.4in]{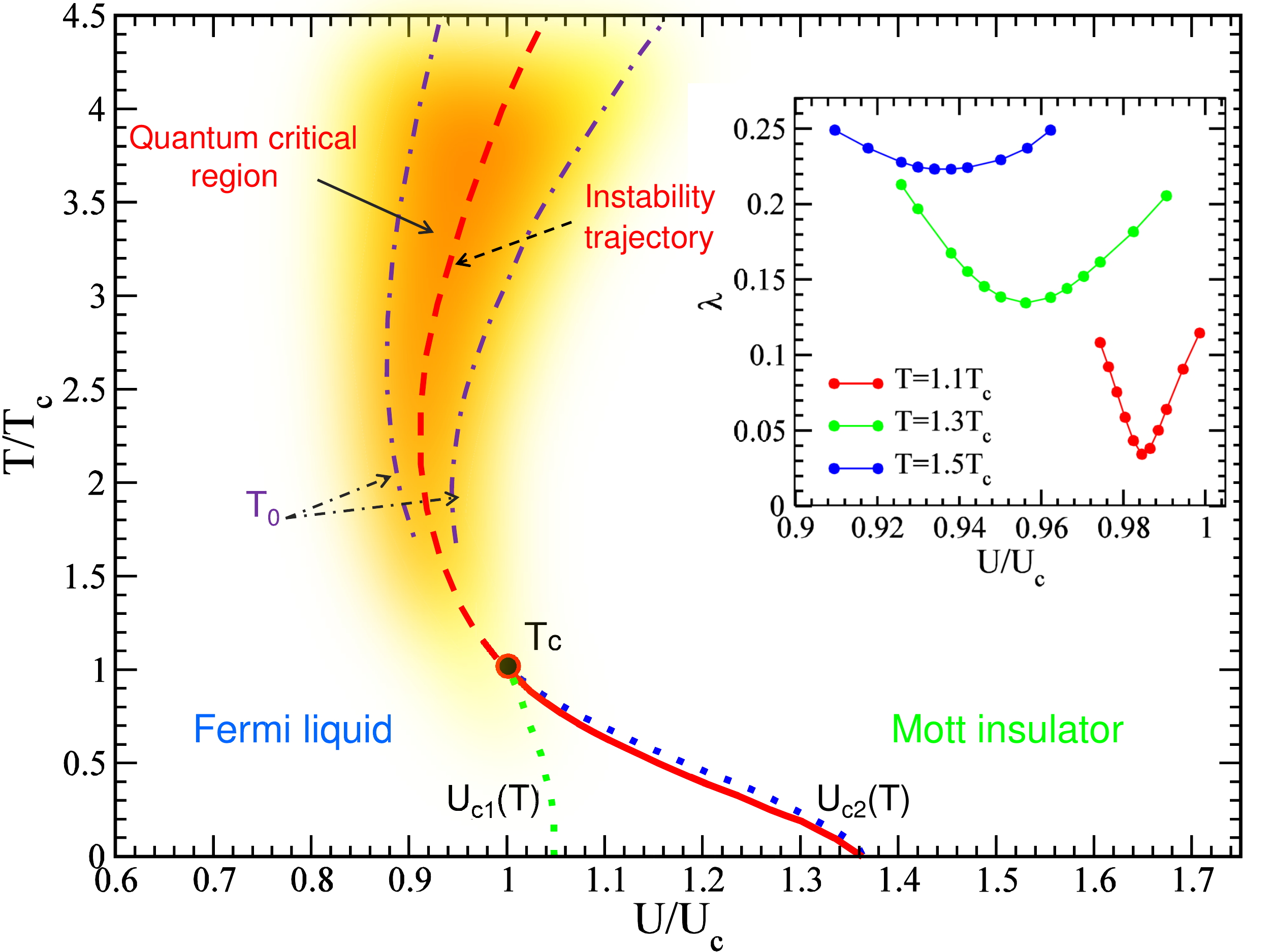} \caption{(Color online) DMFT phase diagram of the fully frustrated half-filled
Hubbard model, with a shaded region showing where quantum critical-like
scaling is found. Metallic $U_{c2}(T)$ and insulating $U_{c1}(T)$
spinodals (dotted lines) are found at $T<T_{c}$; the corresponding
first order phase transition is shown by a thick solid line. The thick
dashed line, which extends at $T>T_{c}$ shows the {}``instability
trajectory'' $U^{\star}(T)$, and the crossover temperature $T_{o}$
delimits the QC region (dash-dotted lines). The inset shows examples
of eigenvalue curves at three different temperatures, with pronounced
minima at $U^{\star}(T)$ determining the instability trajectory.
\vspace*{-12pt}
 }

\end{figure}

It is well known that at very low temperatures $T<T_{c}\sim0.03$,
this model features a first order metal-insulator transition terminating
at the critical end-point $T_{c}$ (Fig. 2), very similar to the familiar
liquid-gas transition \cite{Kotliar2000}. For $T>T_{c}$, however,
different crossover regimes have been tentatively identified \cite{Georges1996,Rozenberg1995}
but they have not been studied in any appreciable detail. The fact
that the first order coexistence region is restricted to such very
low temperatures provides strong motivation to examine the high temperature
crossover region from the perspective of {}``hidden quantum criticality''.
In other words, the presence of a coexistence dome at $T<T_{c}\ll1$,
an effect with very small energy scale, is not likely to influence
the behavior at much higher temperatures $T\gg T_{c}$. In this high
temperature regime smooth crossover is found, which may display behavior
consistent with the presence of a {}``hidden'' quantum critical
(QC) point at $T=0$. To test this idea, we utilize \emph{standard}
scaling methods appropriate for quantum criticality, and compute the
resistivity curves along judiciously chosen trajectories respecting
the symmetries of the problem.

\textit{Instability trajectory formalism}. Previous work has already
recognized \cite{Kotliar2000} that, in order to reveal the proper
scaling behavior close to the critical end-point, one has to follow
a set of trajectories parallel to {}``zero field'' trajectory $U^{*}(T)$.
We thus expect $\delta U\equiv U-U^{\star}(T)$ to play the role of
the scaling variable corresponding to a symmetry-breaking field favoring
one of the two competing (metal vs. insulator) phases. By analogy
\cite{Kotliar2000,Mooij1973} to the familiar liquid-gas transition,
we determine the precise location of such an {}``instability trajectory''
by examining the curvature of the corresponding free energy functional
\cite{Moeller1999}. This curvature vanishes at $T_{c}$ and is finite
and \emph{minimal} at $T>T_{c},$ along this instability line. Consequently,
as in Refs. \cite{Kotliar2000,Moeller1999,Case2007}, our problem
is recast as an eigenvalue analysis of the corresponding free energy
functional $\mathcal{F}[G(iw_{n})]$ for which the DMFT Green's function
solution $G_{DMFT}(iw_{n})$ represents a local extremum, and can
be regarded as a vector in an appropriate Hilbert space.

The free energy near such an extremum can be written as $\mathcal{F}[G(iw_{n})]=\mathcal{F}_{0}+Tt^{2}\sum_{m,n}\delta G(i\omega_{m})M_{mn}\delta G(i\omega_{n})+\dots$,
where \begin{equation}
M_{mn}=\frac{1}{2Tt^{2}}\left.\frac{\partial^{2}{\cal F}[G]}{\partial G(i\omega_{m})\partial G(i\omega_{n})}\right|_{G=G_{DMFT}}\end{equation}
 and $\delta G(i\omega_{n})\equiv G(i\omega_{n})-G_{DMFT}(i\omega_{n})$.
The curvature of the free energy functional is determined by the lowest
eigenvalue $\lambda$ of the \textit{fluctuation matrix} $M$. As
explained in Supplementary Notes \cite{Supplementary}, $\lambda$
can be obtained from the iterative solution of DMFT equations. The
difference of the Green's functions in iterations $n$ and $n+1$
of the DMFT self-consistency loop is given by \begin{equation}
\delta G^{(n+1)}(i\omega_{n})-\delta G^{(n)}(i\omega_{n})=e^{-n\lambda}\delta G^{(0)}(i\omega_{n})\end{equation}
 and therefore $\lambda$ determines the rate of convergence of the
Green function to its solution.

An example of our calculations is shown in the inset of Fig. 2, where
the eigenvalues at several temperatures are plotted as a function
of interaction $U/U_{c}$. The minima of these curves define the locus
of the {}``\emph{instability trajectory}'' $U^{\star}(T)$ , which
terminates at the critical end-point $(U_{c},T_{c})$, as shown on
Fig. 2. Note, that the immediate vicinity $T\approx T_{c}$ of the
critical end-point has been carefully studied theoretically \cite{Kotliar2000},
and even observed in experiments \cite{Limelette2003}, revealing
classical Ising scaling (since one has a finite temperature critical
point) of transport in this regime. In our study, we examine the crossover
behavior at much higher temperatures $T\gg T_{c}$, displaying very
different behavior: precisely what is expected in presence of \emph{quantum
criticality}.

\textit{Resistivity calculation.} To reveal quantum critical scaling,
we calculate the temperature dependence of the resistivity along a
set of trajectories parallel to our instability trajectory (fixed
$\delta U=U-U^{\star}(T)$). Resistivity was calculated using standard
DMFT procedures \cite{Georges1996}, with the maximum entropy method
\cite{Jarrell1996} utilized to analytically continue CTQMC data to
the real axis. The resistivity results are shown in Fig. 1, where
on panel (a) IPT resistivity data for $\delta U=0,\pm0.025,\pm0.05,\pm0.1,\pm0.15,\pm0.2$
in the temperature range of $T\approx0.07-0.2$ are presented (CTQMC
data are not shown for the sake of clarity of the figure). The resistivity
is given in units of $\rho_{Mott}$, maximal resistivity according
to the Boltzmann quasi-classical theory of transport \cite{Hussey2004}.
The family of resistivity curves above ($\delta U>0$) the {}``separatrix''
$\rho_{c}(T)$ (dashed line, corresponding to $\delta U=0$) has an
insulating-like behavior, while metallic dependence is obtained for
$\delta U<0$. %
\begin{figure}[t]
 \includegraphics[width=3.3in]{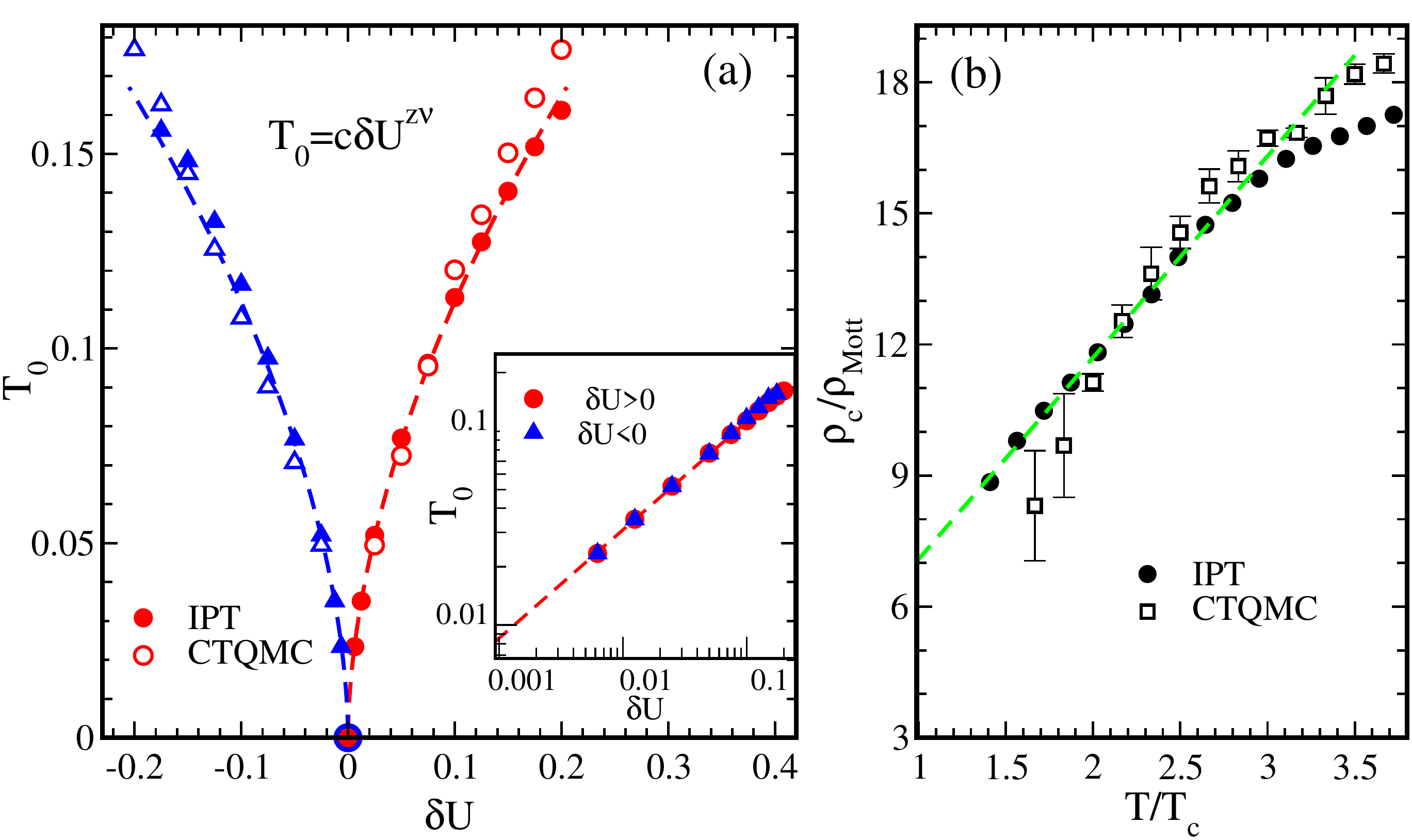} \caption{(Color online) (a) Scaling parameter $T_{0}(\delta U$) as function
of control parameter $\delta U=U-U^{\star}$; the inset illustrates
power-law dependence of scaling parameter $T_{0}=c|\delta U|^{z\nu}$;
(b) Resistivity $\rho_{c}(T)$ of the {}``separatrix''. Excellent
agreement is found between IPT (closed symbols) and CTQMC (open symbols)
results. \vspace*{-12pt}
 }

\end{figure}

\textit{Scaling analysis}. According to what is generally expected
for quantum criticality, our family of curves should satisfy the following
scaling relation:\begin{equation}
\rho(T,\delta U)=\rho_{c}(T)f(T/T_{o}(\delta U)).\label{eq:-3}\end{equation}
 We thus first divide each resistivity curve by the {}``separatrix''
$\rho_{c}(T)=\rho(T,\delta U=0)$ and then rescale the temperature,
for each curve, with an appropriately chosen parameter $T_{0}(\delta U)$
to collapse our data onto two branches, Fig. 1(b). Note that this
unbiased analysis \emph{does not assume} any specific form of $T_{o}(\delta U)$:
it is determined for each curve simply to obtain optimum collapse
of the data \cite{NOTE-InstabilityLine}. This puts us in a position
to perform a stringent test of our scaling hypothesis: true quantum
criticality corresponds to $T_{o}(\delta U)$ which vanishes at $\delta U=0$
and displays power law scaling with the same exponents for both scaling
branches. As seen in Fig.~3(a) $T_{0}$ falls sharply as $U=U^{\star}$
is approached, consistent with the QC scenario, but opposite to what
is expected in a {}``classical'' phase transition. The inset of
Fig.~3(a) with log-log scale shows clearly a power-law behavior of
$T_{0}=c|\delta U|^{z\nu}$, with the estimated power $\left(z\nu\right)_{_{\delta U<0}}^{^{IPT}}=0.56\pm0.01$
for {}``metallic '' side, and $\left(z\nu\right)_{_{\delta U>0}}^{^{IPT}}=0.57\pm0.01$
for an insulating branch.

We also find (Fig.~3(b)) a very unusual form of our critical resistivity
$\rho_{c}(T)$, corresponding to the instability trajectory. Its values
largely exceeds the Mott limit, yet it displays metallic-like but
non-Fermi liquid-like temperature dependence $\rho_{c}(T)\sim T.$
Such puzzling behavior, while inconsistent with any conventional transport
mechanism, has been observed in several strongly correlated materials
close to the Mott transition \cite{Hussey2004,Powell2006}. Our results
thus suggest that it represents a generic feature of Mott quantum
criticality.

\textit{$\beta-$function and mirror symmetry of scaled curves.} To
specify the scaling behavior even more precisely, we compute the corresponding
$\beta$-function \cite{Dobrosavljevic1997} $\beta(g)=\frac{d\ln g}{d\ln T}$,
with $g=\rho_{c}/\rho$ being the inverse resistivity scaling function.
Remarkably (Fig. 4(a)), it displays a nearly linear dependence on
$\ln g$, and is continuous through $\delta U=0$ indicating precisely
the same form of the scaling function on both sides of the transition
- another feature exactly of the form expected for genuine quantum
criticality. This functional form is very natural for the insulating
transport, as it is obtained even for simple activated behavior $\rho(T)$
$\sim e^{-E_{g}/T}$. The fact that the same functional form \emph{persists}
well into the metallic side is a surprise, especially since it covers
almost an order of magnitude for the resistivity ratio. Such a behavior
has been interpreted \cite{Dobrosavljevic1997} to reflect the \textquotedbl{}strong
coupling\textquotedbl{} nature of the critical point, which presumably
is governed by the same physical processes that dominate the insulator.
This points to the fact that our QC behavior has a strong coupling,
i.e. non-perturbative character.

The fact that the $\beta-$function assumes this logarithmic form
on both sides of the transition is mathematically equivalent \cite{Dobrosavljevic1997}
to stating that the two branches of the corresponding scaling functions
display \textquotedbl{}mirror symmetry\textquotedbl{} over the same
resistivity range. Indeed, we find that transport in this QC region
exhibits a surprisingly developed reflection symmetry (dash vertical
lines of Fig. 4(a) mark its boundaries). Such a symmetry is clearly
seen in Fig. 4(b), where the resistivity $\rho/\rho_{c}$ (for $\delta U>0$)
and conductivity $\sigma/\sigma_{c}=\rho_{c}/\rho$ ($\delta U<0$)
can be mapped onto each other by reflection with $\frac{\rho(\delta U)}{\rho_{c}}=\frac{\sigma(-\delta U)}{\sigma_{c}}$
\cite{Simonian1997}. Note that $T/T_{o}=1$ sets the boundary of
the quantum critical region, over which the reflection symmetry of
scaled curves is observed. It is depicted by dash-dotted crossover
lines $T_{0}$ on phase diagram of Fig.~2 \cite{Supplementary}.

These remarkable features of the $\beta-$function, and associated
reflection-symmetry, have been observed earlier in experimental \cite{Abrahams2001,Simonian1997}
and theoretical \cite{Dobrosavljevic1997} studies, which tentatively
associated this with disorder-dominated MITs. Speculation that $\beta\sim\ln g$
reveals disorder as the fundamental driving force for MIT, presumably
reflects the fact that, historically, it has first been recognized
for Anderson transitions \cite{Abrahams1979}. Our work, however,
shows that such behavior can be found even in absence of disorder
- in interaction driven MITs. This finding calls for re-thinking of
basic physical processes that can drive the MIT.

\begin{figure}[!t]
 \includegraphics[width=3.3in]{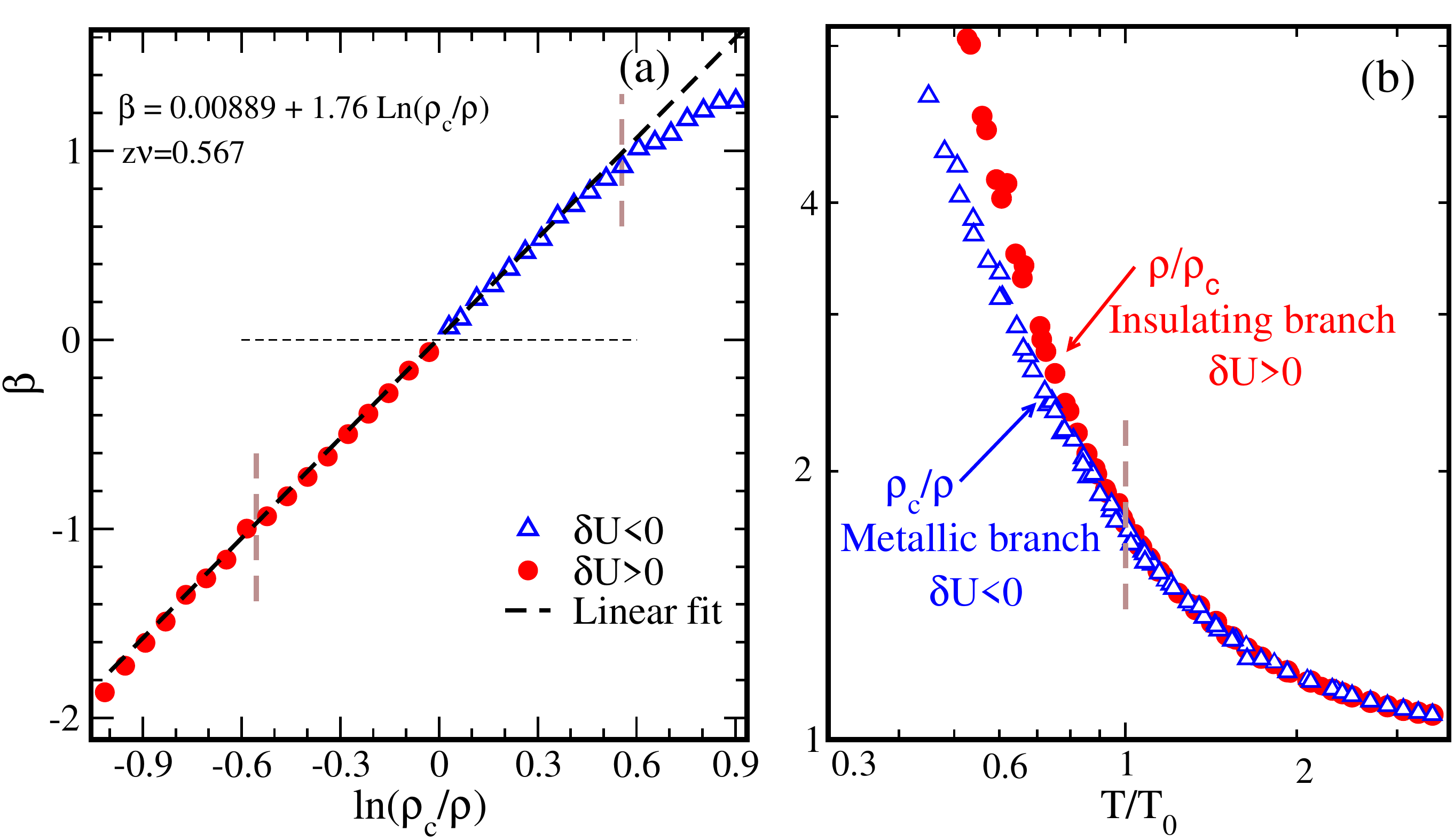} \caption{(Color online) (a) $\beta-$function shows linear in $\ln(\rho_{c}/\rho)$
behavior close to the transition. Open symbols are for metallic branch
($\delta U<0$) and closed ones are for the insulating side ($\delta U>0$)
; vertical dashed lines indicate the region where mirror symmetry
of curves is found. (b) Reflection symmetry of scaled curves close
to the transition.}
\vspace*{-10pt}
 
\end{figure}

\textit{Conclusions:} We have presented a careful and detailed study
of incoherent transport in the high temperature crossover regime above
the critical endpoint $T_{c}$ of a single-band Hubbard model. Our
analysis revealed a so-far overlooked scaling behavior of the resistivity
curves, which we interpreted as evidence of {}``hidden'' Mott quantum
criticality. Precisely locating the proposed QCP in our model is hindered
by presence of the low temperature coexistence dome, which limits
our quantum critical scaling to the region well above $T_{c}.$ Regarding
the nature of transport in the QC regime, we found that the critical
resistivity well exceeds the Mott limit, and yet it - surprisingly
- assumes a metallic form, in dramatic contrast to conventional MIT
scenarios. These features, together with large amounts of entropy
characterizing this entire regime \cite{DeLeo2010}, prove surprisingly
reminiscent of the \textquotedbl{}holographic duality\textquotedbl{}
scenario \cite{Sachdev2010,McGreevy2010} for a yet-unspecified QC
point. Interestingly, the holographic duality picture has - so far
- been discussed mostly in the context of quantum criticality in correlated
metals (e.g. $T=0$ magnetic transitions in heavy fermion compounds).
Ours is the first work proposing that the same physical picture could
apply to quantum criticality found at the MIT.

We believe that our results provide a significant new perspective
on QC around the Mott transition, and a deeper understanding of an
apparent universality in the high temperature crossover regime. Our
method traces a clear avenue for further searches for QC scaling,
which are likely to be found in many other regimes and models.

In particular, it would be interesting to study a corresponding critical
regime by going beyond the single-site DMFT analysis. It was shown
in \cite{Park2008} that inclusion of spatial fluctuations does not
significantly modify the high temperature crossover region in half-filled
Hubbard model. Consequently, we expect our main findings to persist.
An even more stringent test of our ideas should be provided in models
where the critical end-point $T_{c}$ can be significantly reduced.
This may include studies of the Mott transition away from half-fillling
\cite{Sordi2009}, or in systems with frustrations \cite{Camjayi2008,Eckstein-2007}.
In such situations the proposed scaling regime should extend to much
lower temperatures, perhaps revealing more direct evidence of the
- so far - hidden Mott QC point. Our ideas should also be tested by
performing more detailed transport experiments in the relevant incoherent
regime, a task that may be easily accessible in various organic Mott
systems \cite{Kagawa2005}, where $T_{c}$ is sufficiently below room
temperature.

The authors thank K. Haule for the usage of his CTQMC code,
and M. Jarrell for the use of his Maximum Entropy code for analytical
continuation of the CTQMC data. This work was supported by the National
High Magnetic Field Laboratory and the NSF through grants DMR-0542026
and DMR-1005751 (H.T. and V.D.), and Serbian Ministry of Science under
project No. ON171017 (J.V. and D.T.). D.T. acknowledges support from
the NATO Science for Peace and Security Programme grant No. EAP.RIG.983235.
Numerical simulations were run on the AEGIS e-Infrastructure, supported
in part by FP7 projects EGI-InSPIRE, PRACE-1IP and HP-SEE.

\bibliographystyle{apsrev} 





 \clearpage  



\section{Supplementary Notes: Quantum Critical Transport Near the Mott Transition}

\subsection{Eigenvalue analysis of the free energy curvature}

Here we
present in detail the procedure that we use to determine the
minimum curvature of the free energy functional for a given
temperature. For simplicity we concentrate on the Bethe lattice.
The Ginzburg-Landau free energy functional $F(\vec{G})$ in the
Hilbert space of the Matsubara Green's functions $\vec{G}\equiv
G(i\omega_n)$ is given by
\cite{Georges1996,Muller1999,Kotliar2000}
\begin{eqnarray}
 F(\vec{G}) &=& -Tt^2 \vec{G}^2 + F_{imp}(\vec{G}) \nonumber \\
 &=& -Tt^2\sum_n G^2(i\omega_n) + F_{imp}(\vec{G}), \label{eq1}
\end{eqnarray}
where the first term is the energy cost of forming the Weiss field $\vec\Delta=t^2 \vec G$
around a given site, while the second term is the free energy of an electron at this
site in the presence of the Weiss field.

Close to a local extremum $\vec{G_{0}}$, we can Taylor expand
$F(\vec{G})$ in terms of deviation from this point $\delta\vec{G}=\vec{G}-\vec{G}_{0}$:
\begin{eqnarray}
F(\vec{G}) &=& F(\vec{G}_{0})+ Tt^2 \sum_{mn}\delta G(i\omega_{m})M_{mn}\delta
G(i\omega_{n})+ \dots \nonumber \\
&=& F(\vec{G}_{0})+ Tt^2 \, \delta\vec{G}\hat{M}\delta\vec{G} + \dots
\end{eqnarray}
where
\begin{equation}
\left. M_{mn}=\frac{1}{2Tt^2}\frac{\partial^2 F(\vec{G})}
{\partial G(i\omega_n)\partial G(i\omega_m)}\right|_{\vec{G}=\vec{G}_0} .
\end{equation}
We introduce a gradient function
\begin{equation}
\vec{g}(\vec{G})\equiv \frac{1}{2Tt^2}
\frac{\partial F(\vec{G})}{\partial \vec{G}} =\hat{M}{\delta\vec{G}}
\end{equation}
and define an iteration-substitution procedure by
\begin{equation}\label{eq5}
\delta{\vec{G}}^{(n+1)}=
\delta\vec{G}^{(n)}-\vec{g}(\vec{G}^{(n)}),
\end{equation}
which gives the minimum of the free energy as the iteration step
$n\rightarrow \infty$.
In the case of the free energy functional given by
Eq.~(\ref{eq1}), we have
\begin{equation}
\vec{g}(\vec{G})=\vec{G}_{imp}(\vec{G})-\vec{G},
\end{equation}
and in the iterative procedure the Green's function converges to
the minimum of the free energy which is at the same time also a
self-consistent solution of the DMFT equations, given by the
relation $G_{imp}(i\omega_n)=G(i\omega_n)$.
\begin{figure}[t]
\includegraphics  [width=3.2in]{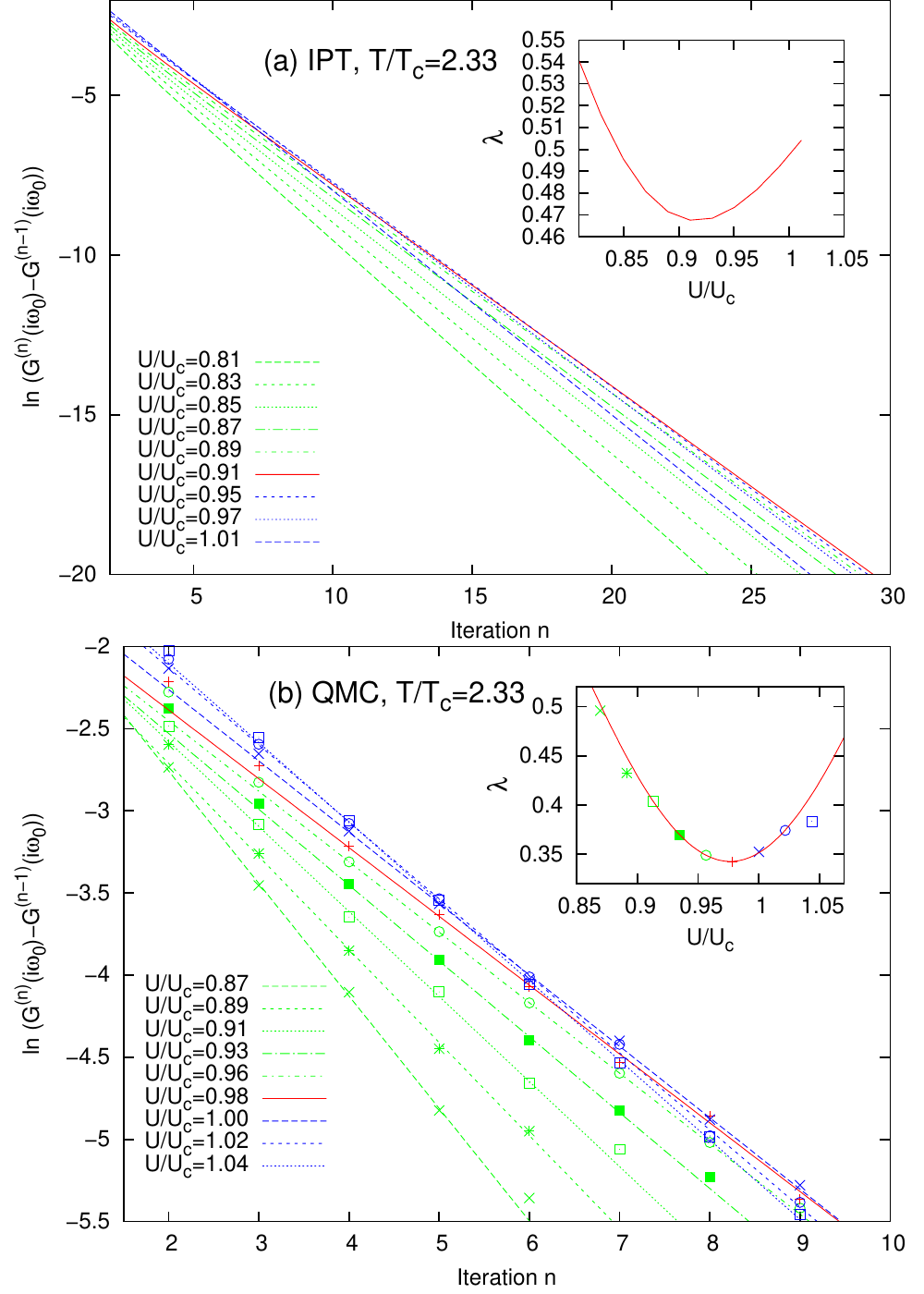}
\caption{Convergence rate in the iterative solution of DMFT
equations at $T/T_c=2.33$ using IPT impurity solver, panel (a),
and CTQMC impurity solver, panel (b). The dashed lines in panel (b)
are linear fits to the data. The insets are the corresponding
eigenvalues determined by the slopes from the main panels.}
\end{figure}

The curvature of the free energy for interaction $U$ and
temperature $T$ can be obtained as follows. The eigenbasis
$\vec{G}_{\alpha}$ and eigenvalues $\lambda_{\alpha}$ of matrix
$\hat M$ are defined by
\begin{equation}
\hat{M}\vec{G}_{\alpha}=\lambda_{\alpha}\vec{G}_{\alpha} .
\end{equation}
We can expand $\delta\vec{G}^{(n)}$ as
\begin{equation}
\delta\vec{G}^{(n)}=\sum_{\alpha}a_{\alpha}^{(n)}\vec{G}_{\alpha}
\end{equation}
where $a_{\alpha}^{(n)}$ are the coefficients of
$\delta\vec{G}^{(n)}$ in the eigenvalue basis. Substituting into
Eq.~(\ref{eq5}), one obtains
\begin{equation}\label{eq9}
\delta\vec{G}^{(n)}=\sum_{\alpha}e^{-nB_{\alpha}}a_{\alpha}^{(0)}\vec{G}_{\alpha} ,
\end{equation}
where
\begin{equation}\label{eq10}
B_{\alpha}= - \ln(1-\lambda_{\alpha}) .
\end{equation}

\begin{figure}[t]
\includegraphics [width=4.2in, angle=270]{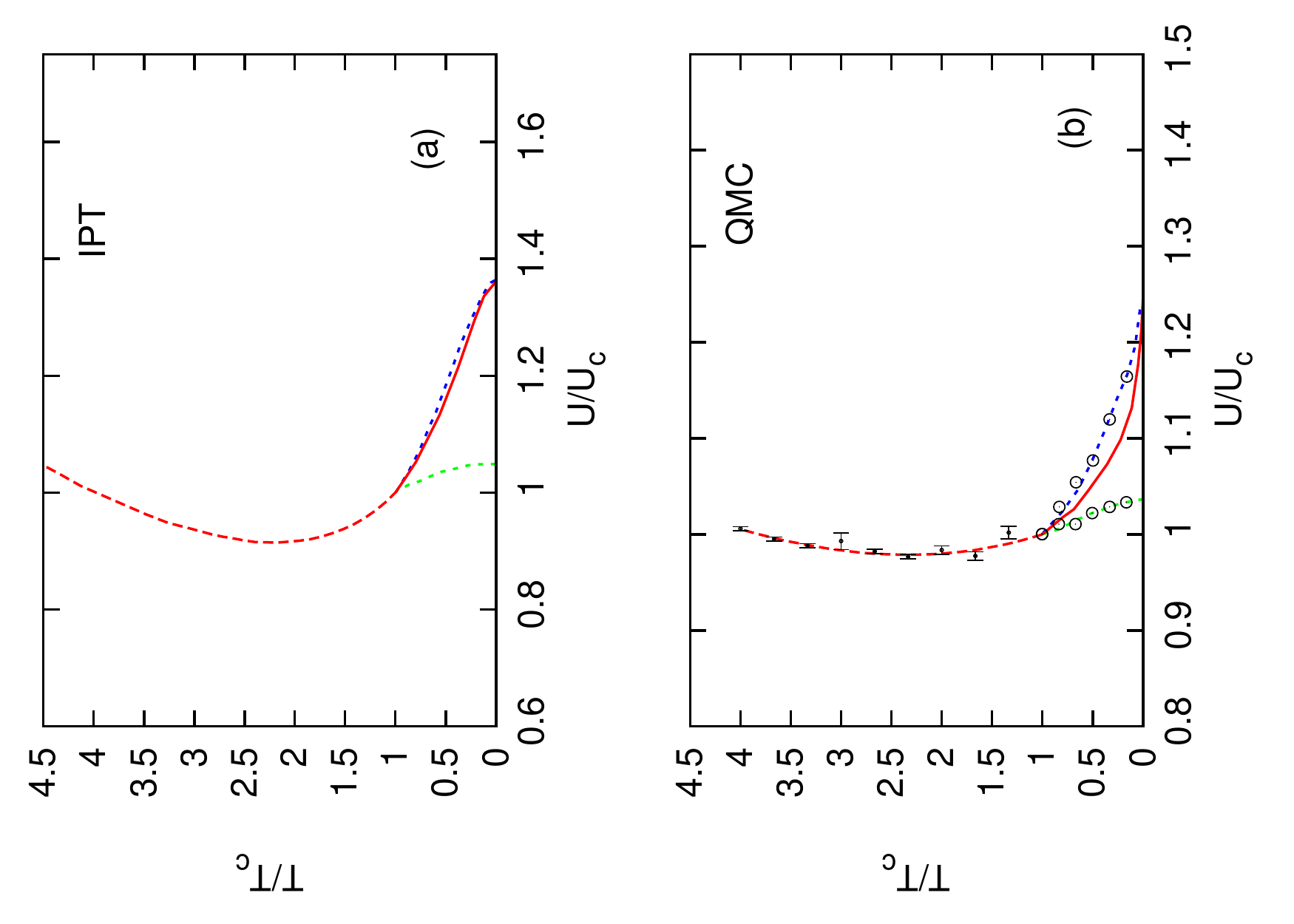}
\caption{Phase diagram obtained with IPT, panel (a), and CTQMC,
panel (b). Temperature and interaction are scaled by their values
at the critical endpoint ($T_c^{IPT}= 0.046$, $T_c^{QMC}= 0.03$
and $U_c^{IPT}= 2.472$, $U_c^{QMC}= 2.3$). Red dashed line is the
instability line $U^*(T)$, full red line is the line of the first
order MIT, and green and blue dotted lines are left and right
spinodals. }
\end{figure}

For large $n$ the term with lowest $B_{\alpha}=B_{\alpha_0}$,
which corresponds to the lowest eigenvalue
$\lambda_{\alpha_0}\equiv \lambda$, is dominant
\begin{equation}
\delta\vec{G}^{(n)}=e^{-nB_{\alpha_0}}a_{\alpha_0}^{(0)}\vec{G}_{\alpha_0},
\, \, \, \, n\gg 1 .
\end{equation}
Here $\alpha_0$ is the coefficient corresponding to the Green
function with the lowest eigenvalue $\lambda$. Now it is obvious
that through iterations, the solution $\vec{G}$ approaches to
$\vec{G}_0$ exponentially along a direction defined by the
eigenvector of $\hat M$ corresponding to its minimal eigenvalue
$\lambda$. The coefficient $B_{\alpha_0}$ and the corresponding
eigenvalue $\lambda$ are then obtained from the slope in the
iterative relation
\begin{equation}\label{eq12}
\ln\left[ G(i\omega_n)^{(n+1)}-G(i\omega_n)^{(n)} \right] =const-nB_{\alpha_0},
\end{equation}
which follows from Eq.~(\ref{eq9}).

In practice, to obtain $\lambda$ (and thus the curvature of free
energy), we monitor DMFT loop convergence rate,
$G(i\omega_o)^{(n+1)}-G(i\omega_o)^{(n)}$, in as many iterations
as possible and then linearly fit
$\ln\left(G(i\omega_o)^{(n+1)}-G(i\omega_o)^{(n)}\right)$ versus
iteration index $n$. Here $\omega_o= \pi T$ is the lowest
Matsubara frequency. For small $\lambda$, $B_{\alpha_0}\approx
\lambda$. We repeat this procedure for different values of $U$ at
the same temperature $T$ to determine $U^*(T)$ in which
$\lambda(U)|_T$ is minimal. It takes few iterations of the DMFT
loop to enter into the linear regime given by Eq.~(\ref{eq12}).

With IPT impurity solver, we can use data from several tens of
iterations to determine the slope $B_{\alpha_0}$, Fig.~1(a). The
solution with CTQMC impurity solver has a statistical error and
the number of iterations is limited before the difference
$|G(i\omega_o)^{(n+1)}-G(i\omega_o)^{(n)}|$ becomes too small and
acquires a large relative statistical error. Nevertheless, we were
able to determine rather precisely the eigenvalue $\lambda$ and
the interaction $U^*(T)$ for which it becomes minimal, Fig.~1(b).
The "instability line" corresponding to the minimum curvature of
the free energy is shown in Fig.~2(a) (IPT phase diagram), and
Fig.~2(b) (CTQMC phase diagram). Error bars in Fig.~2(b) are
estimates of the uncertainty in the position of the instability
line.


\subsection{Details of the scaling procedure}


The resistivity $\rho(T,\delta U)$ is calculated along the lines
parallel to the instability line $U^*(T)$. Here $\delta U =
U-U^*(T)$. The resistivity is fist divided by its value
$\rho_c(T)$ at $\delta U = 0$, Fig.~3(a). Then for each $\delta U$
the temperature axis is scaled by $T_0$ where the scaling
parameter $T_0$ is chosen to collapse the data onto two branches:
insulating-like for $\delta U>0$ and metallic-like for $\delta
U<0$. The scaling was done in such a way that data were collapsed
on the lowest curves with $\delta U=\pm0.025$ as shown in
Fig.~3(b). The scaling parameter $T_0$ has a power law form
$T_{0}=c|\delta U|^{z\nu}$, where the prefactor $c$ depends on
this referent value of $U$.

\begin{figure}[t]
\begin{raggedright}
\includegraphics[width=3.2in]{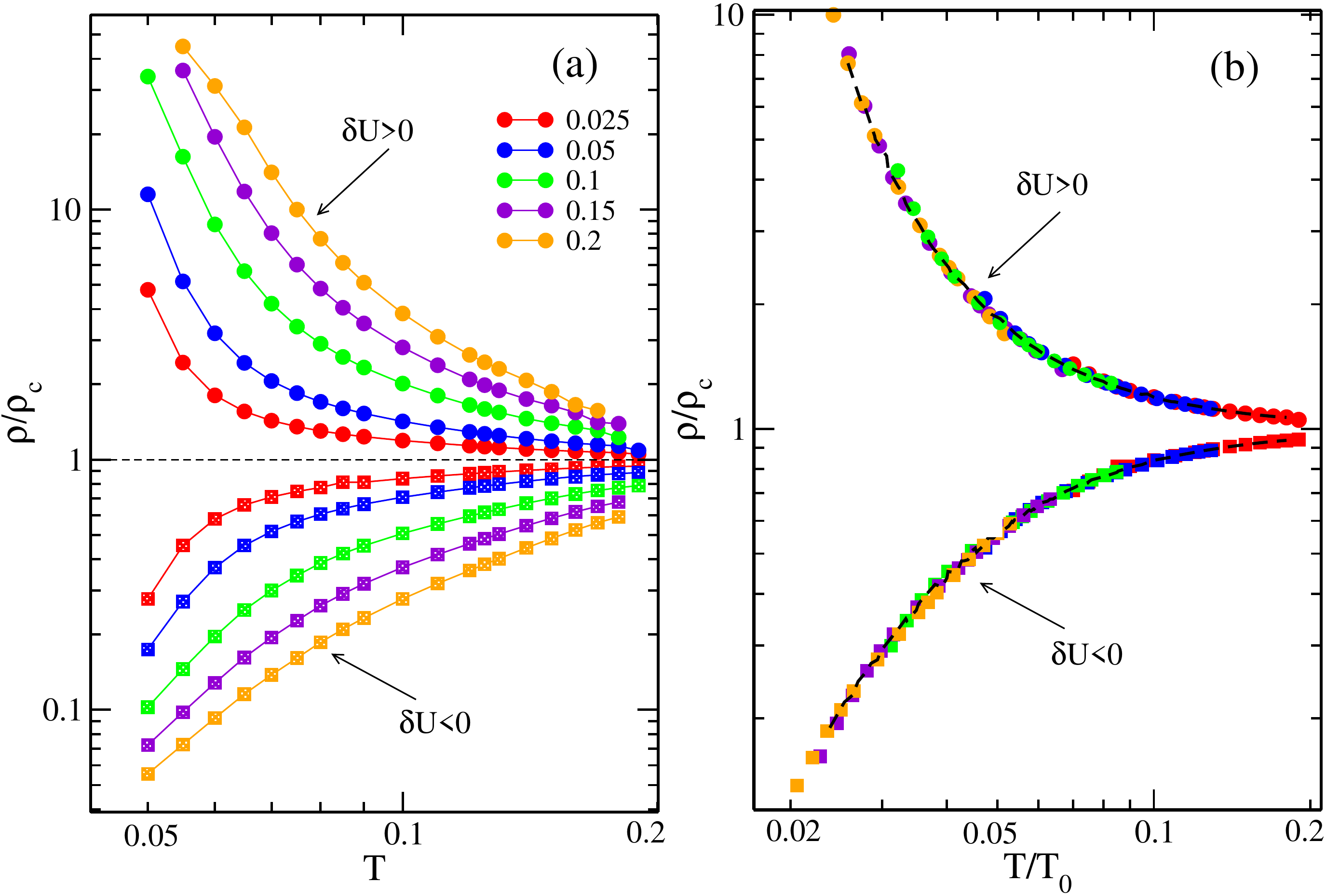}
\par\end{raggedright}
\noindent \caption{(a) Resistivity as function of temperature for
$\delta U=\pm0.025,$0.05,0.1,0.15,0.2. (b) By scaling the data
along $T-$axis by $T_{0}$ data are collapsed onto two branches.
Data are collapsed on the lowest $\delta U=\pm0.025$ curves. }
\end{figure}


Our data exhibit a reflection symmetry which is seen in Fig.~4(a),
where the resistivity $\rho/\rho_{c}$ (for $\delta U>0$) and
conductivity $\sigma/\sigma_{c}=\rho_{c}/\rho$ ($\delta U<0$) can
be mapped onto each other by reflection with $\frac{\rho(\delta
U)}{\rho_{c}} =\frac{\rho_{c}(-\delta U)}{\rho}$.
\begin{figure}[h]
\includegraphics[width=3.2in]{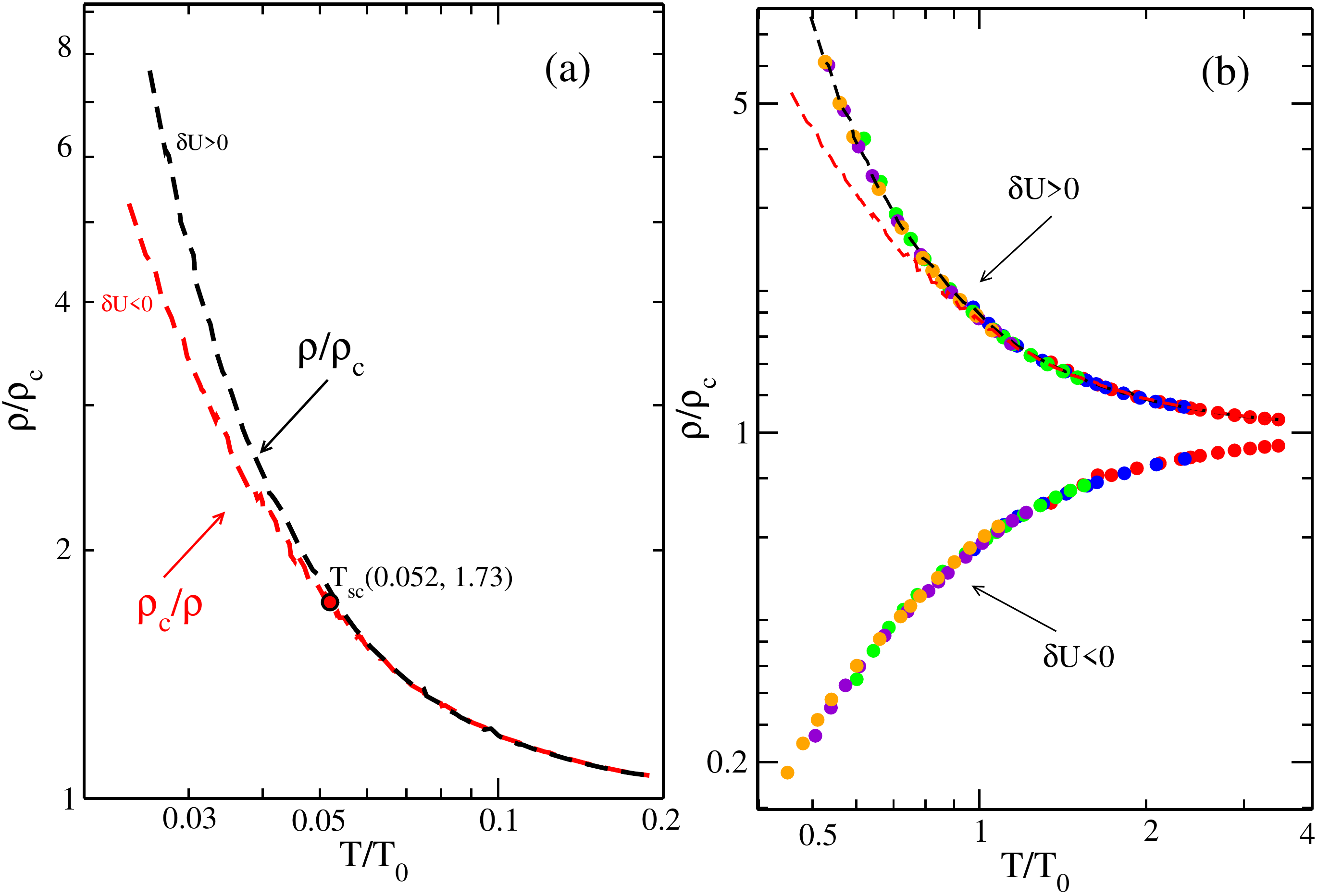}
\caption{(a) Total scaled data (for clarity shown by single
branches) exhibit reflection symmetry for $T\geq T_{sc}$. (b)
$T/T_{0}$-axis is rescaled by $T_{sc}$ in such a way that
$T/T_{o}=1$ set the boundary of quantum critical region over which
the reflection symmetry of scaled curves is observed.}
\end{figure}
\begin{figure}[t]
\includegraphics[width=3.2in]{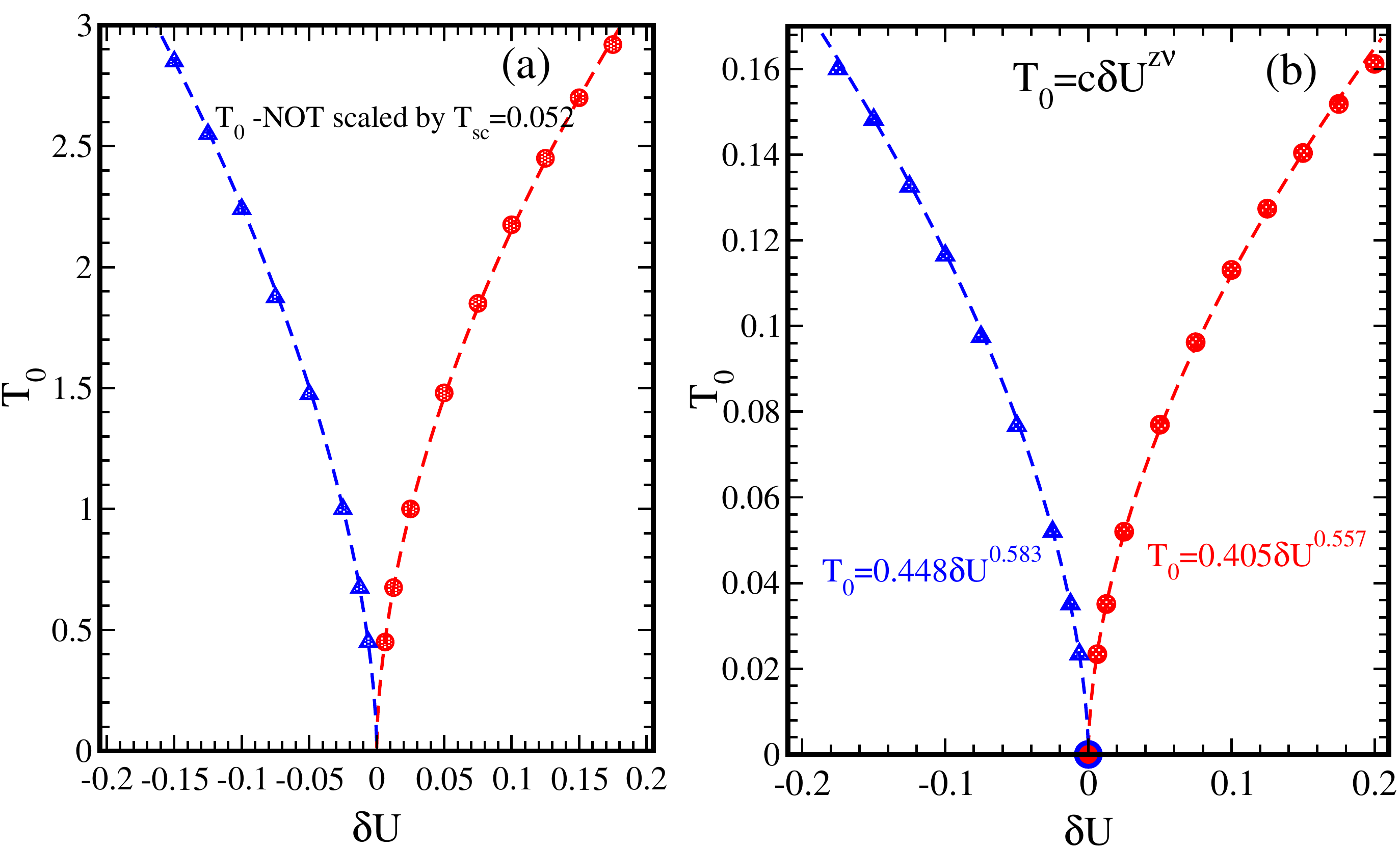}\caption{(a) Scaling temperature $T_{0}$ vs.~$\delta U$ obtained from
scaling procedure shown on Fig.~4(a). (b) $T_{0}$ vs.~$\delta U$
after rescaling by $T_{sc}$. The boundary of the quantum critical
region is now given by condition $T/T_{o}=1$. }
\end{figure}

The standard estimate for the scale (prefactor) of the crossover temperature is obtained by 
requiring that the scaling variable $x=T/T_{0}=1$
at the point where the scaling function changes its functional form; in our case this corresponds 
to the temperature below which the mirror symmetry of the scaling curves no longer holds. Before rescaling the
prefactor, this is found at $x^{*}=T_{sc}/T_{o}=0.052$. Our final
form of scaled data is shown in Fig.~4(b), where $T/T_{0}$ - axis
is rescaled by $x^{*}=T_{sc}/T_{o}$ so that $T/T_{o}=1$ sets the
{\textit {boundary of the quantum critical region}}. The scaling
parameter $T_{0}$ as function of $\delta U$ is shown in Fig.~5(a)
(before rescaling with $x^*$) and in Fig.~5(b) (after rescaling
with $x^*$). The corresponding values for $c$ and $z \nu$ from the
power law fit are also given.

We can now plot the crossover temperature $T_{0}$ setting the
boundary of QC region on our phase diagram. $T=T_0$ condition is
equivalent to
\begin{equation}
T_{0}=c|\delta U(T_{o})|^{\nu
z}=c\delta|U-U^{\star}(T_{o})|^{z\nu}, \label{eq:}\end{equation}
where $U^{\star}(T_{o})$ is value of $U$ at temperature $T=T_{o}$,
along the instability line. This equation implicitly defines the
crossover line $T_{o}(U)$. Alternatively, we can invert this
dependence to describe the same crossover line as $U_{0}(T)$ that
takes the form
\begin{equation}
U_{o}^{\pm}(T)=U^{\star}(T)\pm\left(T/c\right)^{1/z\nu}.\label{eq:-1}\end{equation}
As we can see from this expression, the crossover line approaches
the instability line at low $T$ and diverges away from it at high
$T.$ The phase diagram including the instability line and the
crossover temperature $T_0$ is shown on Fig.~6.


%
%
\begin{figure}[h]
\includegraphics[width=3.3in]{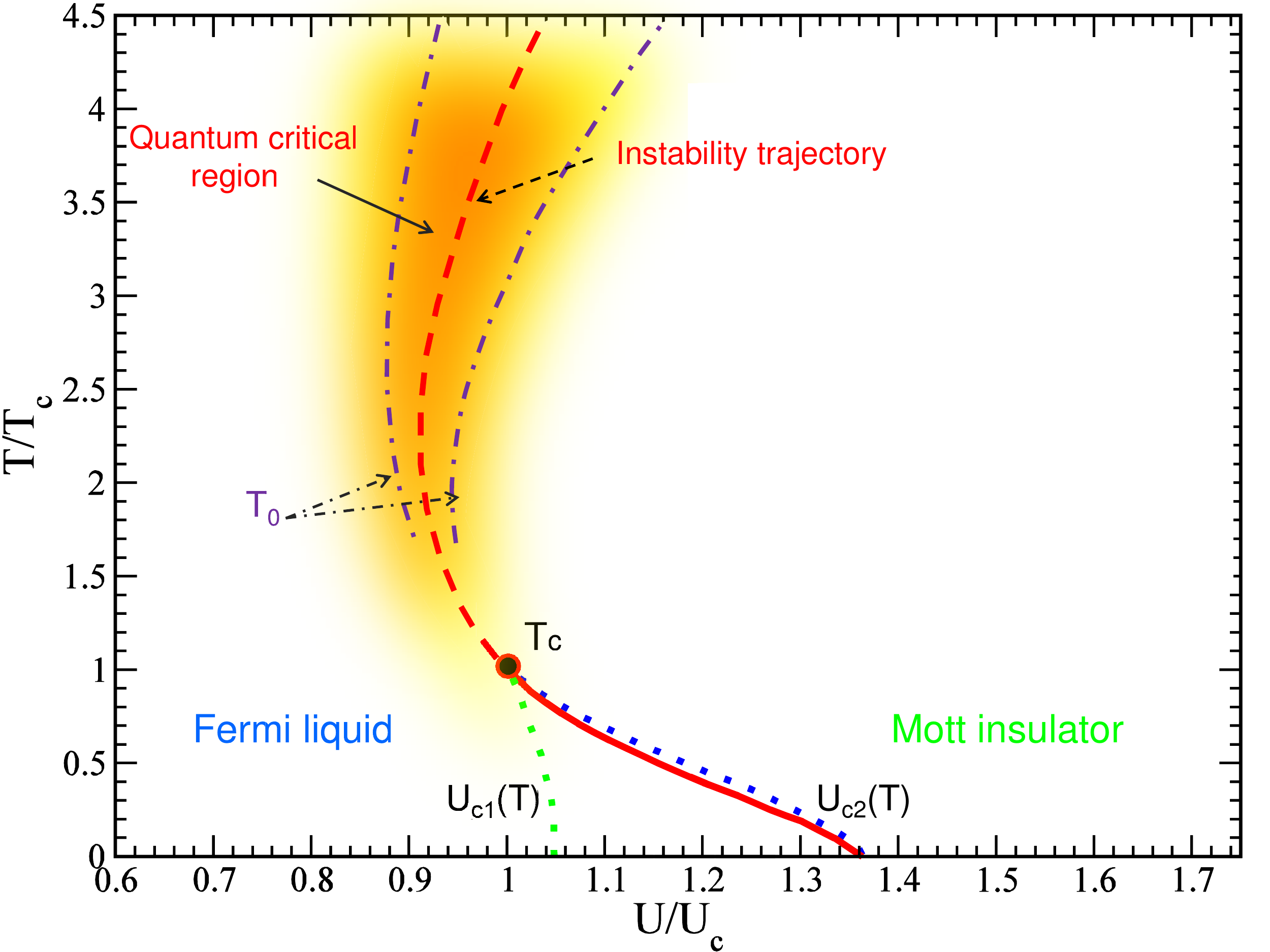}
\caption{(Color online) DMFT phase diagram of the fully frustrated
half-filled Hubbard model. The thick dashed line, which extends at
$T>T_{c}$ shows the {}``instability trajectory'' $U^{*}(T)$, and
the crossover temperature $T_{o}$ delimits the QC region
(dash-dotted lines). \vspace*{-6pt}
 }
\end{figure}

\bibliographystyle{apsrev}


\end{document}